\documentclass{vgtc}                          

\graphicspath{{figures/}{pictures/}{images/}{./}} 

\usepackage{times}                     

\usepackage{tabu}                      
\usepackage{booktabs}                  
\usepackage{lipsum}                    
\usepackage{mwe}                       

\usepackage{mathptmx}                  

\usepackage{enumitem}
\usepackage{comment}

\vgtccategory{Research}

\vgtcinsertpkg

\title{Exploring Eye Tracking to Detect Cognitive Load in Complex Virtual Reality Training}

\author{Mahsa Nasri
\and Mehmet Kosa 
\and Leanne Chukoskie
\and Mohsen Moghaddam 
\and Casper Harteveld 
}
\affiliation{\scriptsize Northeastern University\thanks{e-mail:\{nasri.m, m.kosa, l.chukoskie, m.moghaddam, c.harteveld\}@northeastern.edu}}

\teaser{
  \centering
  \includegraphics[width=1\textwidth]{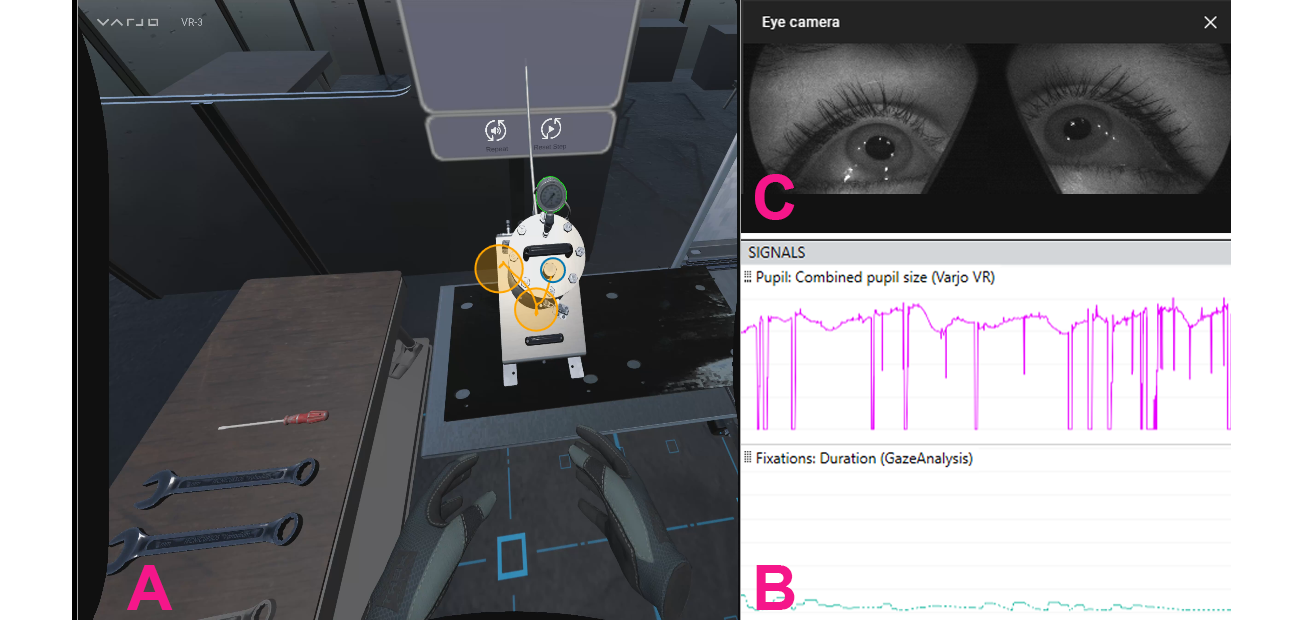}
  \caption{ (A) Processing data in iMotions Lab of a user in the VR training environment interacting with powder feeder; the yellow circles are fixations. (B) The mean pupil dilation signal for both eyes is pink, and the fixation duration signal is the blue color; both signals are processed in iMotions Lab. (C)~User's eyes captured in Varjo VR-3 eye tracking cameras.}
  \label{fig:teaser}
}

\abstract{%

Virtual Reality (VR) has been a beneficial training tool in fields like advanced manufacturing. However, users could experience a high cognitive load due to various factors, such as using VR hardware or tasks within the VR environment. Various studies have shown that eye-tracking has the potential to detect cognitive load, but in the context of VR and complex spatiotemporal tasks (e.g., assembly, disassembly), it is relatively unexplored. Here, we present an ongoing study to detect users' cognitive load using an eye-tracking-based machine learning approach.
We developed a VR training system for cold spray and tested it with 22 participants, obtaining 19 valid eye-tracking datasets and NASA-TLX scores. We applied Multi-Layer Perceptron (MLP) and Random Forest (RF) models to compare the accuracy of predicting cognitive load (i.e., NASA-TLX) with pupil dilation and fixation duration. Our preliminary analysis demonstrates the possibility of using eye tracking to detect CL in complex spatiotemporal VR experiences and motivates our further explorations.
} 


\begin{document}


\firstsection{Introduction}

\maketitle

Virtual Reality (VR) provides an immersive, safe, and credible alternative to real-life training environments \cite{choi2015virtual, bailenson2018experience}. However, most current VR training applications provide unvaried training to all users and do not consider individual differences. One of these key differences is that individuals can experience different Cognitive Load (CL) levels while doing the same task. CL is the mental effort used in working memory \cite{souchet2022measuring}, which can impact learning and performance. A high CL can hinder learning, while an optimal level---neither too frustrating/hard nor boring/easy---can improve it.
A promising solution to consider different cognitive abilities is adapting the training and customizing the content based on the user's CL. Therefore, adaptive training based on the participant's CL can significantly improve training outcomes \cite{10395602, paas2003cognitive}. 
However, detecting high CL in VR environments poses unique challenges. Unlike static environments, dynamic environments include continuous interaction, shifting audio-visual stimuli, and multitasking, all adding to the variation in cognitive demands and making it difficult to accurately capture and analyze CL. In traditional settings, CL can often be assessed using more controlled, static tasks, but these methods fall short in dynamic scenarios where the cognitive demands are constantly shifting \cite{zander2011towards}. Additionally, subjective methods of measuring CL, such as self-reports, may not be practical in VR applications \cite{gao2024exploring,szczepaniak2024predictive}.
A promising solution is using physiological sensors to capture users' cognitive state while they are in the VR. Previous studies have explored CL using various physiological sensors, including heart rate variables, electrodermal activity, electroencephalography, and eye-tracking signals \cite{armougum2019virtual, ahmadi2023cognitive, bodaghi2024multimodal}.
Eye-tracking, in particular, offers a non-intrusive and practical means of measuring CL in VR \cite{souchet2022measuring}. Previous studies have shown correlations between eye-tracking metrics and CL, specifically pupil dilation \cite{souchet2022measuring, thomay2023towards}, but there is limited research on applying these findings in dynamic VR environments \cite{vulpe2023multimodal}.
A vital step towards developing such adaptive training applications is investigating intelligent tools that detect high CL reliably and non-intrusively. Due to complex and non-linear relationships between eye-tracking metrics and CL, Machine Learning (ML) techniques have been under attention in recent years \cite{szczepaniak2024predictive, miles2024cogload}. Current findings reveal that ML models are useful indicators of CL using eye features, obtaining an accuracy of up to 88\% \cite{shojaeizadeh2019detecting, skaramagkas2021cognitive, gao2024exploring}.

\textbf{Problem Statement}
While VR offers a helpful tool for training, it often lacks the consideration of individual cognitive differences. We aim to leverage ML techniques to predict user's cognitive load based on eye-tracking data. This approach aims to create a reliable prediction system that can be integrated into VR training applications in real-time to personalize the training based on the user's cognitive load. This paper demonstrates this possibility by determining if eye-tracking features, particularly pupil dilation and fixation duration, can accurately predict a user's self-reported CL (as measured through NASA-TLX). 

\section{Method}
\begin{figure}[ht!]\label{fig:user}
        \centering
        \includegraphics[width=0.9\linewidth]{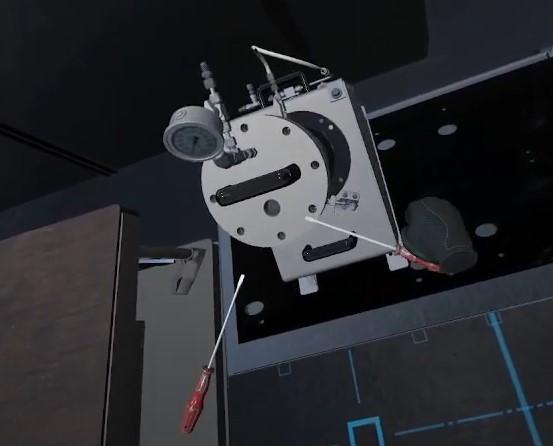}
        \caption{A user while (dis)assembling the powder feeder.}   
\end{figure}

We developed a VR training for \textit{cold spray}, an advanced manufacturing technology that applies coatings of metallic or non-conductive substances to another surface through gas-powered high-velocity spray \cite{papyrin2006cold}. The virtual environment was developed to simulate a cold spray laboratory \cite{nasri2024designing}. First, we designed a VR tutorial room where participants could get familiar with the cold spray environment and interact with objects they would later use during the study. After completing the tutorial, they teleport to the main VR cold spray lab to begin the assembly/disassembly of the Powder Feeder (PF). For context, a PF is a device used to accurately dispense and control the flow of powdered materials in industrial processes, such as in cold spray. A table next to the PF contains all the necessary virtual tools, such as wrenches and screwdrivers. In front of the user, behind the PF, is a panel that displays the task instructions and buttons to request help, reset, or repeat a step (Fig.~\ref{fig:teaser}). The disassembly task consists of 12 steps, where the user uses tools to remove the parts of the PF in the correct order and manner. These two tasks are designed to be completed in 15 minutes, although this may vary depending on the participant. The task descriptions are tailored for novices, so participants with varying expertise in the cold spray process should be able to complete them successfully.
This study focused on evaluating the assembly and disassembly of the \textit{powder feeder}, a primary task in the \textit{cold spray} process (Fig.~\ref{fig:user}).

\subsection{Apparatus and Procedure}
The virtual environment was developed using the Unity engine (version 2020.03.34f1) \cite{unity2024}. We used a VR-3 Varjo headset with a horizontal 115° field of view, 90 Hz refresh rate, and 200 Hz frame rate to capture eye movements, including gaze direction and pupil dilation for each eye.
We recruited 22 participants, all students at the same university with varying levels of familiarity with VR, and recruited them via posters and email ads around the campus. 19 self-identified as male and 3 as female. Gender identification options included were “woman, man, non-binary, prefer to self-describe, and prefer not to say.” Due to technical issues, 3 were discarded, and we had 19 valid eye-tracking datasets. Their age range was 19-28 ($M$ = 23.33, $SD$ = 3.12 ). Our study was approved by the IRB. 
After they had signed a written informed consent form, we showed an introductory video about cold spray to provide context to participants who did not have prior experience with the process. Then, the participants started the VR training. First, a standard 5-point eye-tracking calibration routine was performed in the VR training. Afterward, participants started a tutorial to get familiar with the VR interactions (e.g., grabbing). Next, participants disassembled the \textit{powder feeder} consisting of 12 steps. Then, they moved to the assembly module, which consisted of 11 steps.
After the VR session, participants completed the short version of the NASA Task Load Index (NASA-TLX) to assess the cognitive load of using VRTA \cite{hart2006nasa}. We also collected demographic information at this point. All the surveys were collected via Qualtrics. 

\subsection{Predictors, target variables and Model Architectures}
In this study, our main goal was to explore a binary classification to detect cognitive load. We used Python 3.12.0 with \textit{sklearn}, \textit{TensorFlow}, \textit{Pandas}, \textit{Matplotlib}, and \textit{NumPy} libraries. In pre-processing, we removed the data before starting the VR training tutorial (e.g., calibration and waiting for the app to load). Then, we used the Fast Fourier Transform (FFT) to denoise the signal. Afterward, we normalized the data on a scale of 0 to 1. Previous studies show that pupil dilation is strongly correlated with CL \cite{ahmadi2023cognitive, eckert2021cognitive, lee2023measuring} and longer fixation duration can reflect high CL \cite{liu2022assessing,meghanathan2015fixation}. Hence, we extracted fixation duration and pupil dilation using iMotions Lab software (version 10) \cite{imotions2024} with built-in R Notebooks. Based on established protocols \cite{salvucci2000identifying}, we selected the Velocity-Threshold Identification (I-VT) filter to extract fixations, and the minimum fixation duration was set at 60 ms. Afterward, we calculated the average pupil dilation of both eyes within each fixation and defined fixation duration and pupil dilation as predictors. The target variable was the mental demand subsection of the NASA-TLX, which reveals the mental and cognitive effort required for the task. We analyzed the distributions of the NASA-TLX mental workload subsection and found that the population could be evenly split into low and high groups, with scores of 1 to 4 classified as low and 5 to 7 as high. Accordingly, we labeled participants into either a high or low group and assigned each the corresponding target variable.
Due to the non-linear nature of the predictors and target variables and high-dimensional data, two classifiers—Multi-Layer Perceptron (MLP) and Random Forest (RF)—were trained to predict the target variable. 
We used a sliding window, a common approach in ML \cite{gao2023exploring}, with a size of 2000. 

The MLP model was designed with five hidden layers. We used the hyperbolic tangent activation function across all hidden layers due to its ability to model complex, non-linear relationships. The model was trained using the Adam optimizer with a learning rate of 0.00001, which was selected to ensure stable convergence over 500 epochs. A batch size of 256 was chosen to strike a balance between model convergence speed and predictability. Although dropout layers were considered to mitigate overfitting, the model's performance metrics indicated that the current architecture and settings were sufficient to prevent overfitting. The MLP’s performance was carefully monitored, and the training process was optimized to maintain high precision and recall on both the training and test sets.
The RF model was fine-tuned using \textit{GridSearchCV}, a hyperparameter optimization technique that evaluates all possible combinations of specified parameters to determine the best configuration \cite{aurelien2019hands}. 
The parameters we tuned included the number of trees, maximum tree depth, minimum samples required to split a node, minimum samples required at a leaf node, the number of features considered for splitting, and whether to use bootstrap sampling. We used 3-fold cross-validation to ensure that the selected parameters resulted in a model that predicts unseen data well. 
\section{Preliminary Results}
On average, the participants took around $\approx$ 26 minutes to complete the powder feeder module. Despite a relatively small dataset, we obtained strong results.
The MLP model, evaluated on the test dataset, achieved an accuracy and precision of 0.84, indicating strong prediction capability. The RF model reached an accuracy of 0.72 and a precision of 0.73, suggesting reasonable performance but with some overfitting compared to the MLP model. More details are depicted in Tabel~\ref{tbl1}.

\begin{table} [htb] \label{tbl1}
    \centering
    \caption{The results of predicting cognitive load with MLP and RF models using fixation duration and mean of pupil dilation.}
    \resizebox{\columnwidth}{!}{%
    \begin{tabular}{ | m{4em} | m{6em}| m{6em}  | m{6em}  | m{5em}  |} 
    \hline
     \textbf{Model} & \textbf{Accuracy} & \textbf{Precision} & \textbf{Recall} & \textbf{F1}\\
    \hline 
         MLP & 0.84 & 0.84 & 0.94 & 0.88 \\
     \hline
   RF & 0.72  & 0.73  & 0.90 & 0.81 \\
      \hline
  
    \end{tabular}
   
    }
\end{table}

\section{Discussion and Future Work}
This study investigated the feasibility of predicting cognitive load (CL) in virtual reality (VR) training with complex spatiotemporal tasks to develop a novel adaptive VR training. 
The Multi-Layer Perceptron (MLP) model, with an accuracy of 0.84, performed better than the Random Forest (RF), with an accuracy of 0.72. This indicated that the MLP model better predicts new data, making it suitable for adapting to VR training. While achieving high training metrics, the RF model showed a potential overfitting. These findings underscore the complexity of predicting CL in spatiotemporal tasks in VR, where dynamic scenes and interactive tasks provide more challenges. 

However, several concerns must be addressed before implementing an adaptive training system. Privacy is one the main concerns as eye tracking data can be sensitive and reveal personal information about the user such as age, ethnicity, personality traits, emotional state, and certain measures may even reveal specific mental health conditions \cite{kroger2020does}. Hence, ensuring data security and user privacy is essential.
Additionally, while our current models show promising results, they are limited to pupil dilation and fixation duration. Features such as saccade velocity, saccade amplitude, and blink rate could enhance model accuracy and robustness \cite{zhang2017cognitive, abdurrahman2021effects}  but require more advanced models and computational cost. This study should be considered preliminary work due to its limitations, which are a low number of participants and a limited number of eye features.
In future work, we will explore more complex models, such as Convolutional Neural Networks (CNNs)   \cite{yin2018classification} and personalized models, which combine two or more models. A more complex model could improve accuracy in predicting the user's CL in spatiotemporal tasks in VR training. Our future goal is to develop adaptive VR training that personalizes the training experience in real-time based on the user's cognitive load captured in eye-tracking data.


\acknowledgments{
This material is based on work supported by the National Science Foundation (NSF) grant \#2302838 and the National Center for Manufacturing Sciences (NCMS). Any opinions, findings, or conclusions expressed in this material are those of the authors and do not reflect the views of the NSF or NCMS. The authors further thank the Kostas Research Institute (KRI) at Northeastern University, Naval Sea Systems Command (NAVSEA), VRC Metal Systems, and all the cold spray experts, students, and developers who participated in the design of the AR and VR training systems.}

\bibliographystyle{abbrv-doi}

\bibliography{references}

\begin{thebibliography}{10}

\bibitem{abdurrahman2021effects}
U.~A. Abdurrahman, S.-C. Yeh, Y.~Wong, and L.~Wei.
\newblock Effects of neuro-cognitive load on learning transfer using a virtual reality-based driving system.
\newblock {\em Big Data and Cognitive Computing}, 5(4):54, 2021.

\bibitem{ahmadi2023cognitive}
M.~Ahmadi, S.~W. Michalka, S.~Lenzoni, M.~Ahmadi~Najafabadi, H.~Bai, A.~Sumich, B.~Wuensche, and M.~Billinghurst.
\newblock Cognitive load measurement with physiological sensors in virtual reality during physical activity.
\newblock In {\em Proceedings of the 29th ACM Symposium on Virtual Reality Software and Technology}, pp. 1--11, 2023.

\bibitem{armougum2019virtual}
A.~Armougum, E.~Orriols, A.~Gaston-Bellegarde, C.~Joie-La~Marle, and P.~Piolino.
\newblock Virtual reality: A new method to investigate cognitive load during navigation.
\newblock {\em Journal of Environmental Psychology}, 65:101338, 2019.

\bibitem{aurelien2019hands}
G.~Aur{\'e}lien.
\newblock {\em Hands-on machine learning with Scikit-Learn, Keras, and TensorFlow}.
\newblock o’reilly, 2019.

\bibitem{bailenson2018experience}
J.~Bailenson.
\newblock {\em Experience on demand: What virtual reality is, how it works, and what it can do}.
\newblock WW Norton \& Company, 2018.

\bibitem{bodaghi2024multimodal}
M.~Bodaghi, M.~Hosseini, and R.~Gottumukkala.
\newblock A multimodal intermediate fusion network with manifold learning for stress detection.
\newblock {\em arXiv preprint arXiv:2403.08077}, 2024.

\bibitem{choi2015virtual}
S.~Choi, K.~Jung, and S.~D. Noh.
\newblock Virtual reality applications in manufacturing industries: Past research, present findings, and future directions.
\newblock {\em Concurrent Engineering}, 23(1):40--63, 2015.

\bibitem{eckert2021cognitive}
M.~Eckert, E.~A. Habets, and O.~S. Rummukainen.
\newblock Cognitive load estimation based on pupillometry in virtual reality with uncontrolled scene lighting.
\newblock In {\em 2021 13th international conference on quality of multimedia experience (qomex)}, pp. 73--76. IEEE, 2021.

\bibitem{gao2023exploring}
H.~Gao, L.~Hasenbein, E.~Bozkir, R.~G{\"o}llner, and E.~Kasneci.
\newblock Exploring gender differences in computational thinking learning in a vr classroom: Developing machine learning models using eye-tracking data and explaining the models.
\newblock {\em International Journal of Artificial Intelligence in Education}, 33(4):929--954, 2023.

\bibitem{gao2024exploring}
H.~Gao and E.~Kasneci.
\newblock Exploring eye tracking as a measure for cognitive load detection in vr locomotion.
\newblock In {\em Proceedings of the 2024 Symposium on Eye Tracking Research and Applications}, pp. 1--3, 2024.

\bibitem{hart2006nasa}
S.~G. Hart.
\newblock Nasa-task load index (nasa-tlx); 20 years later.
\newblock In {\em Proceedings of the human factors and ergonomics society annual meeting}, vol.~50, pp. 904--908. Sage publications Sage CA: Los Angeles, CA, 2006.

\bibitem{imotions2024}
iMotions.
\newblock imotions lab.
\newblock \url{https://imotions.com/products/imotions-lab/}, 2024.
\newblock Last accessed: 07/26/2024.

\bibitem{kroger2020does}
J.~L. Kr{\"o}ger, O.~H.-M. Lutz, and F.~M{\"u}ller.
\newblock What does your gaze reveal about you? on the privacy implications of eye tracking.
\newblock In {\em IFIP International Summer School on Privacy and Identity Management}, pp. 226--241. Springer, 2020.

\bibitem{lee2023measuring}
J.~Y. Lee, N.~de~Jong, J.~Donkers, H.~Jarodzka, and J.~J. van Merri{\"e}nboer.
\newblock Measuring cognitive load in virtual reality training via pupillometry.
\newblock {\em IEEE Transactions on Learning Technologies}, 2023.

\bibitem{liu2022assessing}
J.-C. Liu, K.-A. Li, S.-L. Yeh, and S.-Y. Chien.
\newblock Assessing perceptual load and cognitive load by fixation-related information of eye movements.
\newblock {\em Sensors}, 22(3):1187, 2022.

\bibitem{meghanathan2015fixation}
R.~N. M~eghanathan, C.~van Leeuwen, and A.~R. Nikolaev.
\newblock Fixation duration surpasses pupil size as a measure of memory load in free viewing.
\newblock {\em Frontiers in human neuroscience}, 8:1063, 2015.

\bibitem{miles2024cogload}
G.~Miles, M.~Smith, N.~Zook, and W.~Zhang.
\newblock Em-cogload: An investigation into age and cognitive load detection using eye tracking and deep learning.
\newblock {\em Computational and Structural Biotechnology Journal}, 24:264--280, 2024.

\bibitem{nasri2024designing}
M.~Nasri, U.~Narayan, M.~Feyyaz~Sonbudak, A.~Simonson, M.~Chiu, J.~Donati, M.~Sivak, M.~Kosa, and C.~Harteveld.
\newblock Designing a virtual reality training apprenticeship for cold spray advanced manufacturing.
\newblock {\em arXiv e-prints}, pp. arXiv--2411, 2024.

\bibitem{paas2003cognitive}
F.~Paas, A.~Renkl, and J.~Sweller.
\newblock Cognitive load theory and instructional design: Recent developments.
\newblock {\em Educational psychologist}, 38(1):1--4, 2003.

\bibitem{papyrin2006cold}
A.~Papyrin, V.~Kosarev, S.~Klinkov, A.~Alkhimov, and V.~M. Fomin.
\newblock {\em Cold spray technology}.
\newblock Elsevier, 2006.

\bibitem{salvucci2000identifying}
D.~D. Salvucci and J.~H. Goldberg.
\newblock Identifying fixations and saccades in eye-tracking protocols.
\newblock In {\em Proceedings of the 2000 symposium on Eye tracking research \& applications}, pp. 71--78, 2000.

\bibitem{shojaeizadeh2019detecting}
M.~Shojaeizadeh, S.~Djamasbi, R.~C. Paffenroth, and A.~C. Trapp.
\newblock Detecting task demand via an eye tracking machine learning system.
\newblock {\em Decision Support Systems}, 116:91--101, 2019.

\bibitem{skaramagkas2021cognitive}
V.~Skaramagkas, E.~Ktistakis, D.~Manousos, N.~S. Tachos, E.~Kazantzaki, E.~E. Tripoliti, D.~I. Fotiadis, and M.~Tsiknakis.
\newblock Cognitive workload level estimation based on eye tracking: A machine learning approach.
\newblock In {\em 2021 IEEE 21st International Conference on Bioinformatics and Bioengineering (BIBE)}, pp. 1--5. IEEE, 2021.

\bibitem{souchet2022measuring}
A.~D. Souchet, S.~Philippe, D.~Lourdeaux, and L.~Leroy.
\newblock Measuring visual fatigue and cognitive load via eye tracking while learning with virtual reality head-mounted displays: A review.
\newblock {\em International Journal of Human--Computer Interaction}, 38(9):801--824, 2022.

\bibitem{szczepaniak2024predictive}
D.~Szczepaniak, M.~Harvey, and F.~Deligianni.
\newblock Predictive modelling of cognitive workload in vr: An eye-tracking approach.
\newblock In {\em Proceedings of the 2024 Symposium on Eye Tracking Research and Applications}, pp. 1--3, 2024.

\bibitem{unity2024}
U.~Technologies.
\newblock Unity.
\newblock \url{https://unity.com/}, 2024.

\bibitem{10395602}
C.~Thomay, A.~Fermitsch, J.~Fessler, P.~Garatva, B.~Gollan, A.~Katharina~Lietz, M.~Matscheko, and M.~Wagner.
\newblock Towards cognitive load-based decision making in vr training.
\newblock In {\em 2023 IEEE 2nd International Conference on Cognitive Aspects of Virtual Reality (CVR)}, pp. 000023--000028, 2023. doi: {{%
10\hspace{.1pt}\discretionary{.}{%
}{.}\hspace{.4pt}1109\discretionary{/}{%
}{/}CVR58941\hspace{.1pt}\discretionary{.}{%
}{.}\hspace{.4pt}2023\hspace{.1pt}\discretionary{.}{%
}{.}\hspace{.4pt}10395602}}


\bibitem{thomay2023towards}
C.~Thomay, A.~Fermitsch, J.~Fessler, P.~Garatva, B.~Gollan, A.~K. Lietz, M.~Matscheko, and M.~Wagner.
\newblock Towards cognitive load-based decision making in vr training.
\newblock In {\em 2023 IEEE 2nd International Conference on Cognitive Aspects of Virtual Reality (CVR)}, pp. 000023--000028. IEEE, 2023.

\bibitem{vulpe2023multimodal}
A.~Vulpe-Grigorasi.
\newblock Multimodal machine learning for cognitive load based on eye tracking and biosensors.
\newblock In {\em Proceedings of the 2023 Symposium on Eye Tracking Research and Applications}, pp. 1--3, 2023.

\bibitem{yin2018classification}
Y.~Yin, C.~Juan, J.~Chakraborty, and M.~P. McGuire.
\newblock Classification of eye tracking data using a convolutional neural network.
\newblock In {\em 2018 17th IEEE International Conference on Machine Learning and Applications (ICMLA)}, pp. 530--535. IEEE, 2018.

\bibitem{zander2011towards}
T.~O. Zander and C.~Kothe.
\newblock Towards passive brain--computer interfaces: applying brain--computer interface technology to human--machine systems in general.
\newblock {\em Journal of neural engineering}, 8(2):025005, 2011.

\bibitem{zhang2017cognitive}
L.~Zhang, J.~Wade, D.~Bian, J.~Fan, A.~Swanson, A.~Weitlauf, Z.~Warren, and N.~Sarkar.
\newblock Cognitive load measurement in a virtual reality-based driving system for autism intervention.
\newblock {\em IEEE transactions on affective computing}, 8(2):176--189, 2017.

\end{thebibliography}
\end{document}